\documentclass[fleqn,usenatbib]{mnras}

\usepackage{newtxtext,newtxmath}
\usepackage[T1]{fontenc}

\DeclareRobustCommand{\VAN}[3]{#2}
\let\VANthebibliography\thebibliography
\def\thebibliography{\DeclareRobustCommand{\VAN}[3]{##3}\VANthebibliography}

\usepackage{graphicx}	% Including figure files
\usepackage{amsmath}	% Advanced maths commands
\title[ICL on Abell 370 and Abell S1063]{The intracluster light on Frontier Fields clusters Abell 370 and Abell S1063}

\author[N. Oliveira et al.]{
N\'{i}colas O. L. de Oliveira,$^{1}$\thanks{E-mail: nicolasoliveira@on.br}
Yolanda Jim\'{e}nez-Teja,$^{2}$
Renato Dupke$^{1,3,4}$
\\
$^{1}$Observat\'{o}rio Nacional, Rua General Jos\'{e} Cristino, 77 - Bairro Imperial de São Crist\'{o}vão, Rio de Janeiro, 20921-400, Brazil\\
$^{2}$Instituto de Astrofisica de Andaluc\'{i}a (CSIC),
Glorieta de la Astronom\'{i}a s/n, Granada, E-18008, Spain\\
$^{3}$Department of Physics and Astronomy, University of Alabama, Box 870324, Tuscaloosa, AL 35487, USA\\
$^{4}$Department of Astronomy, University of Michigan, 311 West Hall, 1085 South University Ave., Ann Arbor, MI 48109-1107, USA
}

\date{Accepted 2022 February 10. Received 2022 February 10; in original form 2021 August 25}

\pubyear{2021}

\begin{document}
\label{firstpage}
\pagerange{\pageref{firstpage}--\pageref{lastpage}}
\maketitle

\begin{abstract}
We analyzed the contribution of the intracluster light (ICL) to the total luminosity of two massive galaxy clusters observed by the Hubble Space Telescope within the Frontier Fields program, Abell 370 ($z$ $\sim$ 0.375) and Abell S1063 ($z$ $\sim$ 0.348), in order to correlate it with the dynamical stage of these systems. We applied an algorithm based on the Chebyshev-Fourier functions called CICLE, specially developed to disentangle the ICL from the light of galaxies and measure the ICL fraction. We measured the ICL fraction in three broadband optical filters, F435W, F606W, and F814W, without assuming any prior hypothesis about the ICL physical properties or morphology. The results obtained from the ICL fraction vary between $\sim7\%-25\%$, and $\sim3\%-22\%$ for both A370 and AS1063, respectively, which are consistent with theoretical predictions for the total amount of ICL obtained by ICL formation and evolution simulations. We found enhanced ICL fractions in the intermediate filter F606W for both clusters and we suggest that this is due to the presence of an excess of younger/lower-metallicity stars in the ICL compared to the cluster galaxies. We conclude that both Abell 370 and Abell S1063 are merging systems since they exhibit a similar feature as merging CLASH and Frontier Fields clusters sub-sample previously analyzed. We compare these results to the dynamical indicators obtained through different methods and we reinforce the use of ICL as a new and independent method to determine the dynamical state of clusters of galaxies.
\end{abstract}


%%%%%%%%%%%%%%%%%%%%%%%%%%%%%%%%%%%%%%%%%%%%%%%%%%

%%%%%%%%%%%%%%%%% BODY OF PAPER %%%%%%%%%%%%%%%%%%

\section{Introduction}

The space between galaxies in clusters is filled by a diffuse, low surface brightness component known as the intracluster light (ICL), composed of stars that are not bound to any specific galaxy of the system, but are bound by the gravitational potential of the cluster. Since it was first reported by \citet{Zwi51}, remarkable advances have been made in the study of the ICL, mainly due to the availability of deep observations such as those from CLASH \citep[Cluster Lensing and Supernova Survey with Hubble,][]{Post2012} and Frontier Fields \citep[FF,][]{Lotz17} %% and BUFFALO \citep[Beyond Ultra-deep Frontier Fields and Legacy Observations,][]{Stein2020} 
programs. Data obtained from ground-based facilities such as the Subaru Telescope \citep{Miyazaki2002}, the SDSS \citep[Sloan Digital Sky Survey,][]{Gunn1998}, and J-PLUS \citep[Javalambre-Photometric Local Universe Survey,][]{Cenarro19} has also significantly improved ICL detections. Also, in addition to traditional methods, the development of sophisticated techniques has been continuously refining ICL analysis, providing more accurate and precise results \citep[e.g.,] []{Guennou12,Adami13,Jimenez16,Ellien19}.

It is believed that the formation of the ICL is strongly linked to the formation process of the cluster itself. As galaxy clusters are formed by hierarchical accretion, their constituent galaxies interact with each other, first within groups that eventually fall towards the potential of the system and, later, within the cluster's own environment. Over time, several dynamical processes rip out stars from their progenitor galaxies, forming and feeding the population of stars within the intergalactic medium. The ICL is believed to originate from five main processes: a) merging events with the brightest cluster galaxy, or BCG \citep{Murante07,Conroy07}; b) merging with other clusters and groups of galaxies \citep{Mihos04,Rudick06}; c) dynamical friction \citep{Rudick09,Contini14}; d) tidal disruption of dwarf galaxies \citep[e.g.,] []{Rudick09,Melnick12,DeMaio15}; and e) \textit{in situ} formation \citep{Sun07,Puch10}. However, in recent years there has been a growing consensus that tidal disruption of dwarf galaxies and \textit{in situ} formation processes are not relevant contributors to the ICL formation \citep[e.g.,] []{Melnick12,MoTr14,DeMaio18}. Specific features in the stellar populations of the ICL (such as morphology, color, and spatial distribution) can then be used to track which of these formation mechanisms contribute most significantly to their formation, according to the evolutionary moment of the cluster.
Thus, the ICL is fundamental to understand the history of hierarchical formation of the cluster, and, in particular, the BCG \citep[e.g.,] []{Feldmeier04,Gonzalez05,Mihos05,KrickBer07,Rudick11,MoTr14,Edwards2016}. %%As it originates from the interactions that occur in such accretion processes, information about this component (such as ICL fraction and evolution) can indicate which mechanisms contribute most to its formation, the time scale of such processes, and the characteristics of the stellar populations. 
The ICL is also important to understand the metal enrichment of the intracluster gas \citep{SL2005}, to determine the precise baryonic fraction in clusters to constrain cosmological parameters \citep[e.g.,][]{Allen04,LinMohr04}, and to account for the excess of UV background light in low-redshift clusters \citep{Welch20}.

The first theoretical studies about ICL formation indicated that the contribution of this component to the total luminosity of the cluster (the so-called ICL fraction, defined as the ratio between the ICL and the total luminosity of the cluster, which we will refer to as ICLf from now on) corresponds to values within a range of 10\% to 55\% \citep{Rich76,Merrit83,Miller83,RichMal83}. These studies did not make any special distinction between the ICL and the BCG, and were based only on the analysis of tidal interactions and/or simulations of individual galaxies orbiting the gravitational potential of the cluster. The initial estimations of the ICLf did not take into account the effects of interactions between galaxies and intermediate-scale structures proper of dense environments, such as remnants of infalling groups, resulting in inaccurate ICL measurements.
However, more recent studies and simulations estimate that the ICLf has an upper limit of 40\% for low-redshift clusters \citep{Rudick11,Contini14}. Many numerical simulations of ICL formation and evolution also predict that the ICLf growth in clusters is directly associated with merger events between massive galaxies and with infalling groups, rising drastically in major interactions \citep{Rudick06,Murante07,Contini14}. The ICLf in clusters that do not experience major mergers events was found to rise more slowly and mildly.

%% ICL can also provide relevant information for understanding the enrichment of metal in the intracluster medium \citep{MoTr18} and restrict precisely the fraction of baryons for use in cosmological simulations \citep[e.g.,][]{Lima03,Allen04,LinMohr04}.

More recently, \citet{Jimenez18} measured the ICLf in eleven massive clusters observed by the Hubble Space Telescope (HST) in order to verify possible relations between the dynamics of these systems and the ICL. The clusters were selected in order to have similar observational characteristics and well-defined dynamical stages. All of them were observed by CLASH and Frontier Fields programs, with the exception of the Bullet cluster. This sample was located at intermediate redshifts, between 0.18 $<$ $z$ $<$ 0.55, and consisted of five relaxed and six merging clusters. By analyzing the ICLf in three distinct optical filters -- F435W, F606W, and F814W -- for each cluster, they discovered the existence of a characteristic feature in the ICLf distribution that appeared only in the merging systems. In the redshift range of the sample, this feature was characterized by an ICLf excess flux in the F606W filter in the merging clusters in the sample, while the ICLf distribution remained approximately flat, within the error bars, for the relaxed ones. %%This peak in the intermediate filter suggests a greater presence of younger and/or of lower-metallicity stars in the ICL than in the galaxies that are members of the cluster, since F606W mainly captures bluer stellar populations, such as A and F-type stars in that redshift range \citep{Morishita17}. 
This excess is consistent with an enhancement of A and F stars at the redshift of the clusters \citep{Morishita17}. These results tied, for the first time, the dynamical stage of galaxy clusters with the ICL. The aim of the present work is to expand this study, adding the analysis of the ICL fraction in the remaining two clusters of the FF program that were not analyzed in \citet{Jimenez18}, Abell 370 and Abell S1063 (also known as RXC J2248.7-4431). We used the CICLE algorithm \citep{Jimenez16}, a mathematical tool developed to disentangle ICL from the light of the galaxies in the cluster, in three broadband filters, F435W, F606W, and F814W, which is the same methodology applied in \citet{Jimenez18}.

The paper is organized as follows: in Section 2 we describe both FF clusters and their observational characteristics. In Section 3 we describe the adopted methodology, detailing the basics of CICLE functioning and the analysis steps. In Section 4 and 5 we present and discuss our results, and in section 6 we draw the conclusions. Throughout the paper, we adopt a standard $\Lambda$CDM cosmology with $H_0$ = 70 km s$^{-1}$ Mpc$^{-1}$, $\Omega_m$ = 0.3, and $\Omega_\Lambda$ = 0.7.

\section{Data}

The main scientific goal of the Hubble Frontier Fields program (ID: 13495, PI: J. Lotz) is to explore the Universe at high redshift, searching for low-surface brightness galaxies at z $>$ 5 by combining observing strategies such as deep multiband HST imaging and strong gravitational lensing of six massive clusters \citep{Lotz17}. The six observed clusters through over 560 HST orbits -- Abell 2744, MACSJ0416.1-2403, MACSJ0717.5 + 3745, MACSJ1149.5 + 2223, Abell S1063, and Abell 370 -- are located in a redshift range of 0.3 $\leq$ $z$ $\leq$ 0.55 and have masses in the order of $\sim$10$^{15}$ $M_\odot$. All of these systems were selected due to the fact that they are strong gravitational lenses. Other characteristics such as the darkness of the sky, galactic extinction, availability of parallel field observations, and accessibility to auxiliary data were considered in the field selection. All clusters were observed with the HST Advanced Camera for Surveys (ACS/WFC) and the Wide Field Camera 3 (WFC3/IR). The observations of ACS/WFC were used in the broadband filters F435W, F606W, and F814W, and those of WFC3/IR in filters F105W, F125W, F140W, and F160W, between 0.4-1.6$\mu$m. Our work focuses on the FF clusters Abell 370 and Abell S1063 (A370 and AS1063, hereafter) using F435W, F606W, and F814W filters.

A370 is a rich galaxy cluster located at $z$ $\sim$ 0.375 \citep{StRo1999}. It was one of the first galaxy clusters with gravitational lenses to be observed, a fact that made A370 one of the most studied systems to date \citep[e.g.,][]{Kneib1993,Smail1996,Mede10,Umestu11}. The virial mass of A370 has been obtained consistently from different X-ray observations, the Sunyaev-Zel'dovich effect, and gravitational lensing, and has a value of $\sim$1 $\times$ 10$^{15}$ M$_\odot$ \citep{MoEtMos07,Richard10,Umestu11}. The radial profiles of ages and metallicities of A370 were obtained using photometry and SED fitting by \citet{MoTr18}. According to the authors, the age of the cluster's central region and the outer regions of the ICL is $\sim$5 Gyr and $\sim$1.5 Gyr, respectively, and the metallicity of the ICL presents a negative radial gradient ranging from [Fe/H] $\sim$0.22 to --0.4.

AS1063 is located at z $\sim$ 0.348 \citep{Gomez12} and it is the most relaxed of the FF clusters and one of the least powerful lenses of the sample \citep{Lotz17}. Its mass, derived by the Sunyaev-Zel'dovich effect, is $M_{500}$ $\sim$ 1.4 x 10$^{15}$ $M_\odot$  \citep{Will11}. The age and metallicity radial profiles of its ICL vary between $\sim$7 Gyr in the center of the cluster and $\sim$1.7 Gyr in the outer regions, and between [Fe/H] = 0.21 and $\sim$ --0.3 \citep{MoTr18}.

According to \citet{Morishita17}, the stellar mass in the ICL present in both clusters is $\sim10\%$ of the total stellar components in these systems and it is dominated by moderately old stellar populations ($\sim$1--3 Gyr). The authors also reported that prominent color gradients indicate that at least $\sim5\%-10\%$ of the ICL mass at $R < 150$ kpc is composed of A and earlier-type stars that were possibly stripped from star-forming/infalling galaxies.

\begin{figure*}
  \centering
  \includegraphics[width=1.\textwidth]{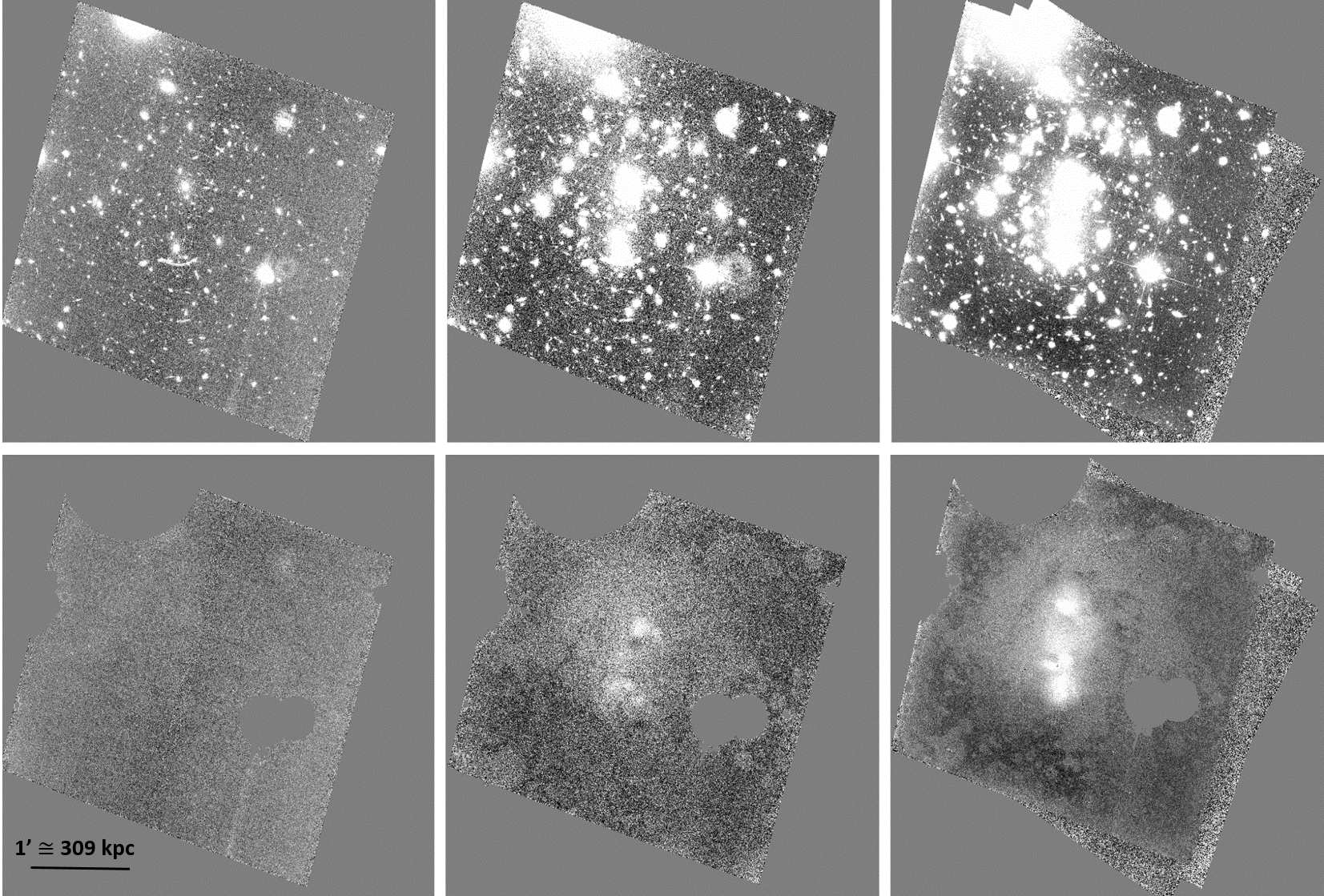} 
  \caption{\textit{Top}: Original images of Abell 370 in the three filters analyzed, F435W, F606W and F814W, from left to right.
  \textit{Bottom}: The ICL map in the same three filters processed by CICLE. All images are in the same scale.}
  \label{fig:a370} 
\end{figure*}

\section{Methodology}

\begin{figure*}
  \centering
  \includegraphics[width=1.\textwidth]{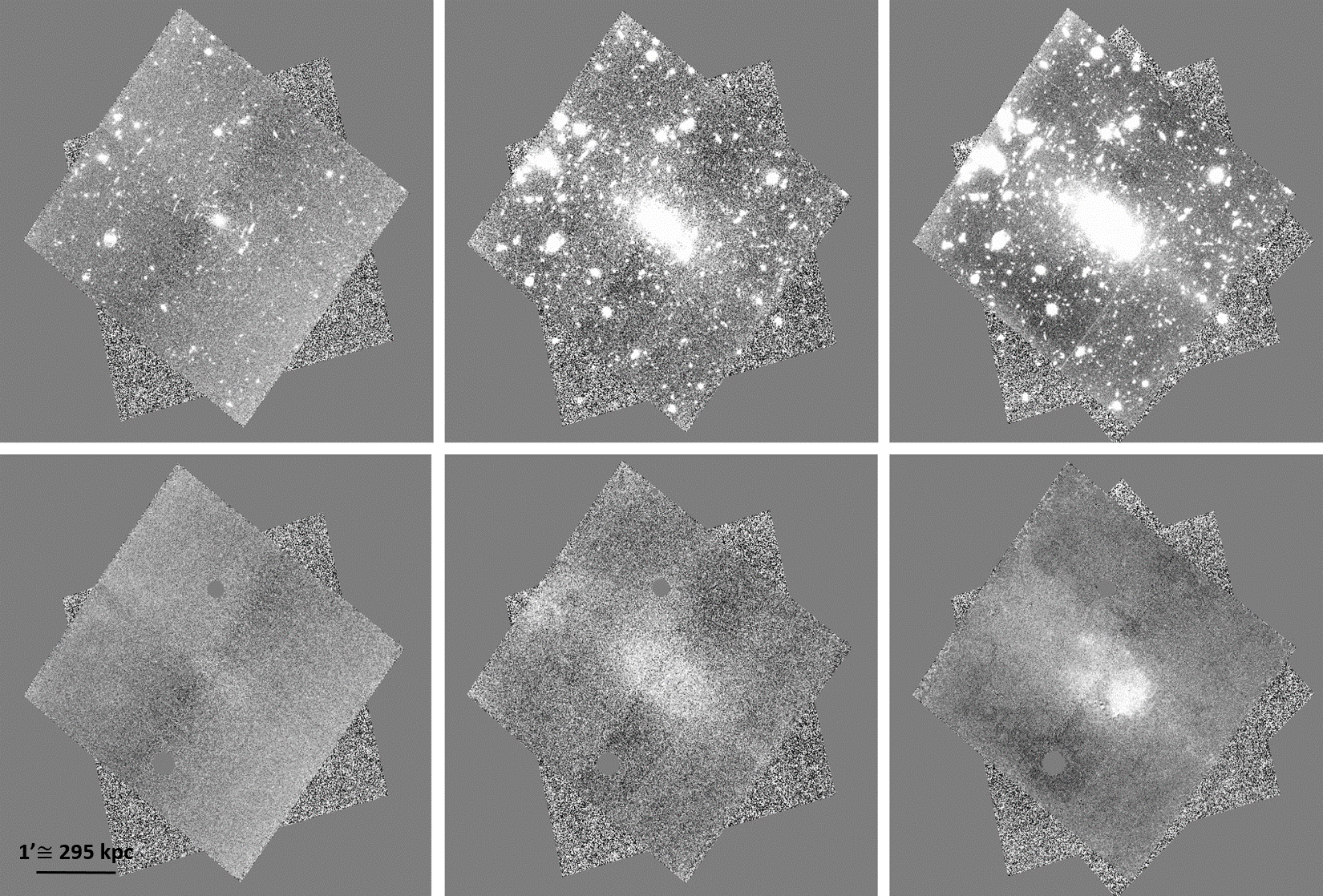} \caption{\textit{Top}: Original images of Abell S1063 in the three filters analyzed, F435W, F606W and F814W, from left to right.
  \textit{Bottom}: The ICL map in the same three filters processed by CICLE. All images are in the same scale.}
  \label{fig:as1063} 
\end{figure*}

To analyze the ICL in these two clusters, we used an algorithm based on the so-called Chebyshev-Fourier functions \citep[CHEFS,][]{Jimenez12} known as CICLE \citep[CHEFs Intracluster Light Estimator,][]{Jimenez16} designed to disentangle the ICL from the total light distribution of the galaxies. It is a mathematical tool that estimates the ICLf without the need of any prior hypothesis on the physical or morphological properties of ICL.

CHEFs are orthonormal bases built in polar coordinates combining Chebyshev rational functions and Fourier series, resulting in highly flexible functions characterized by the remarkable feature of modeling a wide variety of light distributions, including those with extended halos. CICLE combines the modeling power of CHEFs with mathematical tools for characterizing images, such as curvature parameters, and multi-resolution decomposition of data (i.e., morphological transforms and wavelets), to accurately separate and analyze the ICL.

Since CHEF functions constitute a mathematical basis, by definition, they are capable of fitting any sufficiently smooth distribution. Saturated stars are outside that range since their diffraction spikes are sharp, placing them out of the space of mathematical functions capable of being fitted with the CHEFs. So, to begin the analysis of ICL in the selected clusters, we masked out all the brightest stars in each image. %% excluding these regions in the final measurement of the ICLf.

We used SExtractor \citep{BerAr1996,Bertin11} to detect and identify the objects in the images, and the modeling was properly done by CHEFs. Since BCGs have their own extended haloes, the disentanglement of their light component from the ``true'' ICL requires particular attention. For them we performed a specific analysis based on a parameter called Minimum Principal Curvature \citep[MPC,][]{PatriMae10}, to correctly identify the points where the transition between the BCG and the ICL occurs.

Once the light from all galaxies is removed, the remaining image will contain only the ICL and the instrumental background. Although it would be ideal to use observations of parallel fields to estimate the background (as in \citet{Jimenez16}), the available nearby observations with the same detectors were not made at the same epoch as those of our clusters and, therefore, they cannot be used as reference. To subtract the background we used SExtractor, with a specific parameter configuration chosen to minimize misidentification of the ICL with the sky noise that was previously tested in \citet{Jimenez18}. %%...which interpolates a set of discrete values of background estimated by the pixels of the regions that do not contain any objects. The area of these regions is defined by the parameter BACK\_SIZE, which has typical values of 16 or 32, to ensure that small fluctuations on small scales are considered in the estimate. Another necessary parameter for estimating background is BACK\_FILTERSIZE, related to the size of the filter used to smooth the discrete values of the background before interpolating them.

Then, we built an image that had only the cluster members and the ICL by using spectroscopic redshifts. Although it is also possible to obtain the cluster membership from photometric redshifts, we choose not to use them to improve data accuracy and reliability. We thus applied two different methods to identify the cluster members from the spectroscopic information: the PEAK and shifting gapper methods \citep{Owers2011}. The PEAK method roughly discards galaxies by studying their distributions in redshift. In a simplified way, it consists of identifying the peak in a histogram of redshifts, which is usually located around the cluster average, and then identifying the galaxies surrounding this peak as possible members of the system. The shifting gapper method is used to refine the rough selection made in the previous technique by analyzing the peculiar velocities of galaxies in relation to the cluster radius.

Once the cluster members were identified, we re-added the CHEF models of the galaxy members to the ICL image to obtain an image that contains only the luminous contribution from the cluster. Then we measured the ICL fraction by calculating the flux obtained in increasingly larger areas, both for the ICL map and for the total light map, which is the sum of the ICL and the light from the galaxy members of the cluster. For the central region, we used the ICL natural contours (i.e., ICL isocontours/isophotes) as the standard area and then we used ellipses for the external regions. We applied the same contours to both maps, the ICL and total light maps, and built radial flux profiles to calculate the ICLf. The radius up to which the ICLf can be measured is defined by the point at which the ICL flux is minimal. This point is usually easily identifiable by the CICLE algorithm because the profile exhibits an artificial ascending behaviour at larger radii, due to the effects of accumulation of light at the edges of the images or from other instrumental sources of error \citep{Jimenez16}. Finally, we estimated the ICLf corresponding errors for each filter through statistical comparisons with simulated images. %%We will present these processes in more detail in the following section.

For further information and technical details about CICLE algorithm and CHEFs mathematical background, we refer the reader to \citet{Jimenez12} and \citet{Jimenez16}.

\section{Results}

We estimated the ICLf of Abell 370 and Abell S1063 in three ACS/WFC broad bands filters, F435W, F606W, and F814W. Figures \ref{fig:a370} and \ref{fig:as1063} present a comparison between the original images in the three filters for both A370 and AS1063, and their corresponding ICL maps, all matched to the same scale. 

To determine which galaxies are members of A370, we used the catalog of spectroscopic redshifts by \citet{Lagattuta19}, the largest so far, which contains 584 redshifts. Within these objects, we identified 242 as cluster members. For AS1063 we used a catalog that combined redshifts from the CLASH-VLT Large Programme and VLT/MUSE spectroscopy \citep[Rosati et al. in prep,][]{Karman15,Caminha16,Karman17}, which contains 285 spectroscopic redshifts, where we found 148 objects to be cluster members. 
%% Figure 3 presents an example of the final contours of the ICL used for measuring the central region flux and the radial flux profiles of F814W in AS1063.

\begin{figure*}
  \centering
  \includegraphics[width=1.\textwidth]{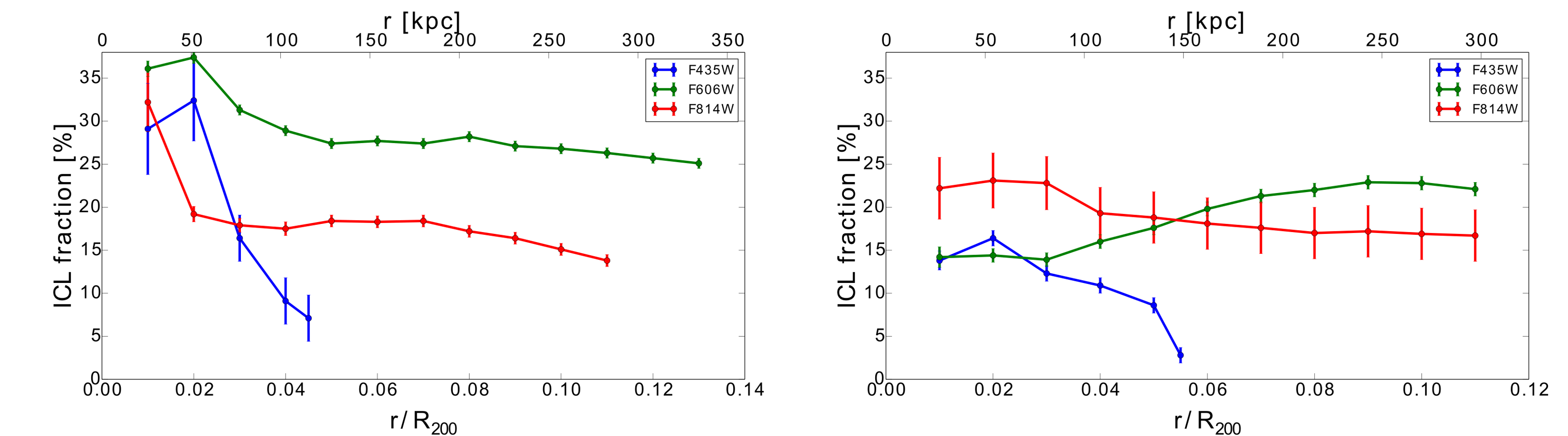} \caption{ICL fractions measured for F435W, F606W, and F814W filters as a function of $R_{200}$ for A370 (left) and AS1063 (right). For A370 we adopted $R_{200} = 2.57$ Mpc \citep{lah09} and for AS1063 we adopted $R_{200} = 2.7$ Mpc \citep{Sartoris20}.}
  \label{fig:iclfr200} 
\end{figure*}

\begin{table*}
\centering
    \caption{ICL fractions and ICL radius measured for A370 and AS1063.}
    \begin{tabular}{lcccccc} 
    \hline
    & \multicolumn{2}{c}{F435W} & \multicolumn{2}{c}{F606W} & \multicolumn{2}{c}{F814W}\\  
    \cline{2-7} 
    Cluster & ICLf [\%] & R [kpc] & ICLf [\%] & R [kpc] & ICLf [\%] & R [kpc]\\ 
    \hline 
    Abell 370 & 7.1 $\pm$ 2.7 & 114.2 & 25.1 $\pm$ 0.4 & 331.1 & 13.8 $\pm$ 0.7 & 291.3\\ 
    Abell S1063 & 2.8 $\pm$ 1.0 & 149.2 & 22.1 $\pm$ 0.8 & 300.6 & 16.7 $\pm$ 3.1 & 295.9\\ 
    \hline
    \end{tabular}
    \label{tab:1}
\end{table*}

The errors associated with the ICLf measurements come from three distinct independent sources. The first one is the photometric error, inherent to the quality of the observed data, which is determined from the observational characteristics of the data, such as the gain, rms, area, and the flux from all cluster members and the ICL. The second source of error is associated with the cluster
membership estimation, which accounts for the underestimation of the total luminosity due to the use of spectroscopic redshifts. We used the limiting magnitudes of the spectroscopic catalogs to calculate the absolute magnitude in which our spectroscopic sample is complete. Then, we used the luminosity functions calculated by \citet{Mede10} for A370 and by \citet{Connor17} for AS1063 to estimate how our measurements are affected by the use of spectroscopic redshifts. The third source of error is inherent to the CICLE method, due fundamentally to the difficulty and intrinsic imprecision in the mathematical process of separating the BCG from ICL. This error is estimated empirically, based on mock images constructed with the same observational characteristics as the original data. CICLE simulates 10 images for each one of the three filters, so that they contain the same geometric and physical information as the original ones. In other words, we simulated an image with a BCG and ICL that have the same geometry and fluxes as the real structures. We generated a composite surface built with three exponential profiles with the same central positions, effective radii, and surface brightness profiles as the BCGs and the ICL from the original data. Then, we added ten realizations of noise (so that the final image has the same signal-to-noise ratio as the original observations), producing ten randomly different images. The final empirical error is the quadratic mean of the estimated individual errors. Finally, the final total error is obtained from the addition in quadrature of the photometric, cluster membership, and empirical errors. 

We  summarize the resulting ICL fractions for each filter and  the corresponding radii for both clusters in  Table  1. Although we use total ICL fractions to estimate the dynamical state of the clusters (that is, the total flux of the ICL, up to its limit of detection), we plot in Figure \ref{fig:iclfr200} the ICL fractions measured as a function of $R_{200}$ (the radius at which the galaxy density is 200 times the critical density) to facilitate the comparison with other works. For A370, we adopted $R_{200} = 2.57$ Mpc derived from the cluster velocity dispersion of 1263 km s$^{-1}$ \citep{lah09}, and for AS1063 we adopted $R_{200} = 2.7$ Mpc derived from the velocity dispersion profile of the BCG and the velocity distribution of cluster members \citep{Sartoris20}.

We discuss these results in the next section, comparing them with those found in the literature and explaining how they are related to the dynamical stage of clusters. 

\section{Discussion}

Here we discuss and compare our results with two previous works in the literature that study the ICL in Frontier Fields clusters \citep{Morishita17,MoTr18}. It should be noted that although \citet{MoTr19} analyzed the ICL in all six FF clusters, they did so for an entirely different purpose. The authors compared the spatial distribution of the ICL with the bi-dimensional distribution of dark matter given by gravitational lensing models. They found promising correspondence between them, showing that the ICL follows the total dark matter distribution in all clusters, which suggests that ICL is a good tracer for the distribution of the underlying dark matter. It also should be noted that neither A370 nor AS103 were included in the samples chosen by \citet{MoTr14} or \citet{Jimenez18}.

The work by \citet{Morishita17} analyzed the ICL in the FF clusters through a method of ICL reconstruction that relies on galaxy fitting. The BCGs do not receive different or special analysis and are fitted with Sérsic profiles which are applied to regions of fixed sizes of 300 $\times$ 300 pixels. A constant sky level in each of these regions was identified as the local ICL and the weighted mean of these overlapping regions as the total ICL. Although the ICL fraction is not analyzed in that work, the authors calculated values for the ICL mass fraction (defined as the stellar mass in the ICL compared to the total stellar mass in the cluster) by SED fitting the broadband photometry of the ICL in different regions, finding values between a range of $\sim7\%-23\%$ for radii less than 300 kpc and between $\sim4\%-19\%$ for radii less than 500 kpc. Specifically, they found ICL mass fractions of $\sim$15$\%$ ($R\leq300$ kpc) and $\sim$12$\%$ ($R\leq500$ kpc) for A370, and $\sim$23$\%$ ($R\leq300$ kpc) and $\sim$19$\%$ ($R\leq500$ kpc) for AS1063.

The work by \citet{MoTr18} analyzed the properties of the ICL stellar populations based on the assumption of a surface brightness limit for galaxies, and that the ICL extends beyond a radius of 50 kpc in relation to the center of the clusters. They found that the ICL generally has overall subsolar metalicities and that the stars are between $\sim$2--6 Gyr younger than those hosted by the most massive galaxies in the system. In order to measure the ICLf, the authors assume a threshold value of $\mu_V$ = 26 mag/arcsec$^2$, finding values in a range of $\sim$1$\%$--4$\%$ for V-band in all six FF clusters. They measured ICL fractions of 1.02$\%$ $\pm$ 0.03$\%$ for A370, and 3.24$\%$ $\pm$ 0.06$\%$ for AS1063. To include the BCG dominated regions that combine both luminous contributions (BCG+ICL, with the radial distance assumed as $R$ $<$ 50 kpc), the authors made a linear interpolation of the ICL surface brightness profiles, finding new values in the range of $\sim$4.8$\%$--13$\%$ in $R$ $<$ $R_{500}$ (where $R_{500}$ is defined as the radial distance where the mean mass density surpass the critical density by a factor of 500) for both clusters, finding 4.8$\%$ $\pm$ 1.7$\%$ for A370, and 13.1$\%$ $\pm$ 2.8$\%$ for AS1063. It is important to note that the ICL, when measured within different radii, produces completely different values of the ICLf. In fact, the authors analyzed the ICL up to a radius of $\sim$120 kpc, which is comparable only to the ICL radius that we obtained for the F435W filter in both clusters (see Table \ref{tab:1}).

The different techniques used for the ICL analysis in these two works result in different values for the ICLf, which make a detailed comparison between their results and ours unfeasible, due to the lack of common parameters to compare them. This is a common issue when comparing different ICL works. Different analysis methods, filters, and assumptions made in determining the ICLf and other ICL physical properties measurements make a fair comparison between them extremely difficult if not impossible \citep{Rudick11}. For example, different assumptions about the surface brightness threshold significantly affect the measurements of the ICLf, leading to different values. Furthermore, the surface brightness threshold presents a problem that depend also on the cluster’s evolutionary stage: for example, it underestimates the ICLf at the exact moment of the merger, since the ICL surface brightness tends to increase and becomes more concentrated as major mergers occur \citep{Rudick06}. In general, our values of the ICLf for A370 and AS1063 are relatively higher than those reported by \citet{MoTr18}. One of the possible explanations for this is the fact that we examine the specific contribution of ICL in BCG dominated areas by using a fully mathematical analysis, not making any assumption on the morphological and physical features of both ICL and BCG components. %% This was not done in the methods used in the works described above in which the ICL present in the central regions is not accurately included in the estimate of the total amount of ICL, since the pixels in these areas are excluded by the pre-defined limits. 
As it is well-known, the ICL is mostly concentrated in the regions %% where the gravitational potential is stronger, therefore,
close to BCGs \citep{Rudick06}, then, the loss of flux in these regions could make the ICL measurements to be underestimated.

\begin{figure*}
  \centering
  \includegraphics[width=1.\textwidth]{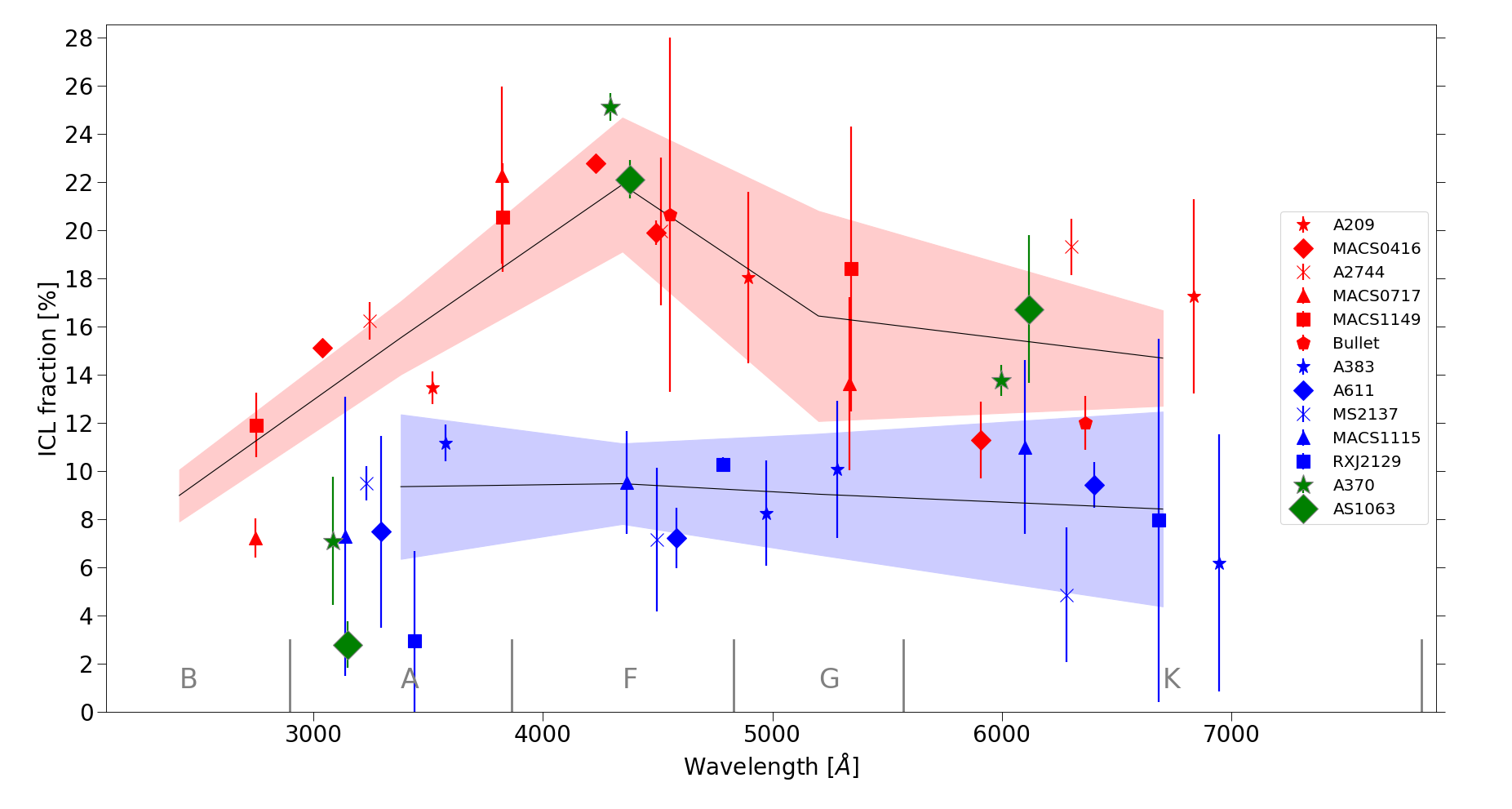} 
  \caption{ICLf measured in the three broad-band filters of A370 (green stars) and AS1063 (green diamonds) added to the sample of clusters analyzed by Jiménez-Teja et al. 2018 at rest-frame wavelength. Merging systems are represented in red and relaxed ones in blue. The black lines indicates the error weighted mean for both merging and relaxed systems, and the red and blue shaded regions represent the mean of the errors.}
  \label{fig:allclusters} 
\end{figure*}

We also compare our measurements with other ICLf analysis made for other cluster samples with similar characteristics to ours. \citet{Giallongo14} analyzed the ICL in the central core of the massive cluster CL0024+17 at $z\sim0.4$ observed with the Large Binocular Telescope. They removed the halo contribution of the galaxies present in the cluster core through profile fitting and derived an ICL profile in the central region. The authors estimated an integrated ICLf in the R-band of $\sim23\%$ within 150 kpc. The R-band at $z \sim 0.4$ is similar to the F606W filter in the redshift of our clusters, therefore these results are comparable to our ICLf values for A370 ($\sim27\%$), and for AS1063 ($\sim18\%$) within 150 kpc (see Figure \ref{fig:iclfr200}). \citet{Burke15} analyzed the ICL in a subsample of 23 massive clusters from CLASH over the redshift range of 0.18 $<$ $z$ $<$ 0.9. They estimated the ICLf of 13 clusters at redshift $z < 0.4$ by assuming a surface brightness threshold of $\mu_B$ = 25 mag/arcsec$^2$ and reported values ranging between $\sim2\%-23\%$. The ICLf in clusters at the redshift interval of $0.3 < z < 0.4$ were between $\sim2\%-8\%$, which are comparable to our ICLf measurements in the F435W filter. %% \citet{MartinezBlake17} analyzed the ICL in the massive cluster Abell 1689 at $z\sim0.18$ with HST data. They estimated the ICLf in the F814W filter by modeling the radial profile of the isophotes using a combination of a core-Sérsic model for the BCG and a more extended standard Sérsic for the ICL, and reported a value of $\sim11\%$. If Abell 1689 is a merging cluster \citep[e.g.,] []{Andersson04}, the ICLf is comparable to our measurements and to those found by \citet{Jimenez18}, although it is in the lower limit of the values found for merging clusters in the F814W band. If it is relaxed \citep[e.g.,] []{Xue08}, the results are even more consistent with ours. 
\citet{Furnell21} analyzed the ICLf in a sample of 18 galaxy clusters detected in XMM Cluster Survey (XCS) data within a redshift range of 0.1 to 0.5. The authors also used the traditional technique of defining the ICL as the component below an isophotal threshold of $\mu_B$ = 25 mag/arcsec$^2$ within an aperture of R$_{X,500}$ (X-ray estimate of R$_{500}$) centered on the BCG. They reported ICLf values within a range of $\sim10\%$ to $\sim38\%$ with an average of $\sim24\%$. Although the upper values are significantly higher than all of our measurements, the average value of the ICLf is comparable to our measurements for the F606W filter.

\citet{Jimenez18} studied the ICL in eleven massive clusters observed by the HST, six from the CLASH program, the Bullet cluster, and four clusters from the FF program: Abell 2744, MACSJ0416.1-2403, MACSJ0717.5+3745, MACSJ1149.5+2223. As mentioned previously, the authors found that the ICLf distribution is a potential indicator of the dynamical stage of the clusters, since merging clusters showed an ICLf excess at a characteristic rest-frame wavelength near 4200{\AA}, while relaxed clusters present nearly constant ICLf measurements with small fluctuations between the three filters. This is due to the fact that within the redshift range of their sample, the F606W filter is the one that best captures the emission peak of type A and F stars. Those type of stars comprises $\sim5\%-10\%$ of ICL mass in FF clusters \citep{Morishita17}. The authors suggest that the excess in F606W ICLf may be caused by the presence of these stars that could have been pulled out of the star-forming/infalling galaxies outskirts %%(or even total disruption of dwarf galaxies) 
during recent mergers.

A particularly interesting feature that our results corroborate is that in the work by \citet{Morishita17} the authors also reported that while all six FF clusters present ICL colors dominated by old stellar population, the non-negligible contribution from the young population is less than 1\% for AS1063. This fraction is greater for the other 5 FF clusters, ranging from $\sim5\%$ to 10$\%$. This result is seen in the left panel of Figure 7 of \citet{Morishita17}, in which it is presented the color-color diagram showing the stellar mass-weighted density map of ICL colors for all FF clusters. Our values of $\sim3\%$ in the ICLf measurement of F435W in AS1063 reinforce that this cluster has a smaller contribution from young populations compared to the other FF clusters through an entirely different method. 

Figure \ref{fig:iclfr200} presents the ICLf profiles measured in the three filters as a function of $R_{200}$ for both clusters. This plot facilitates the comparison of the ICLf fractions between the different filters and also the comparison with systems analysed in other works. A370 has a steeper ICLf profile in small radii and tends to decrease more rapidly than AS1063, which has a smoother decrease. This feature suggests that the center of A370 has a higher amount of the ICL than that of AS1063. This can be due to the fact that A370 has two clear BCGs that could be creating an area of strong tidal forces between them that can strip a lot of material to the ICL in this central region. AS1063 instead has apparently only one BCG, which could prevent a violent strip of stars in the central region, allowing a more spread out distribution of the ICL.

Figure \ref{fig:allclusters} presents the plot exhibited in Figure 1 of \citet{Jimenez18} where we also add the new results obtained by the clusters analyzed here at rest-frame wavelength for comparison. Our ICLf measurements for both clusters, A370 and AS1063 (represented in green stars and diamonds, respectively), exhibit a behavior consistent with the merging clusters of the sample. 

For the case of A370, several different studies in the literature indicate that it is a merging cluster. %as evidenced by the observations of tidal features between galaxies and groups \citep{Oemler97}, distribution of dark matter, gas distribution in X-rays, and velocity distribution \citep{DeFilippis05,Richard10}, mass distribution maps with multiple cores and substructures \citep{Mohammed16},  and, more recently, by hydrodynamical simulations \citep{Molnar20} and by the presence of diffuse synchrotron radio emission \citep{Xie20}
\citet{Oemler97} identified several structures originated from gravitational interactions within A370 that suggest that this system is undergoing successive mergers and tidal interactions. \citet{DeFilippis05} developed a method to constrain A370 intrinsic three-dimensional shape combining both Sunyaev-Zel'dovich effect and X-ray observations and identified a bi-modal structure indicating that it is undergoing a merger of two or more substructures. \citet{Richard10} combined the mass distribution obtained through strong gravitational lensing with observations of X-ray luminosity and cluster's velocity dispersion to conclude that two equally massive groups are merging along the line of sight of A370. \citet{Mohammed16} arrived at a similar conclusion through the identification of multiple cores with interacting clumps of mass in HFF data. A recent work by \citet{Molnar20} used hydrodynamical simulations constrained by optical, X-ray, gravitational lensing, and Sunyaev–Zel’dovich effect observations to infer that A370 is a post major merger system observed after the second core passage, and before the third core passage.

Even though AS1063 is the most ``relaxed"  by comparison with the other FF clusters \citep{Lotz17}, several studies indicate that AS1063 is a merging cluster, which has undergone recent processes of merging galaxies and groups \citep[e.g.,][] {Andersson06,Gomez12,Mohammed16,Xie20}, also using observations of radio emission structures, luminosity and gas temperature in X-rays, and interacting structures seen in optical. Through the analysis of velocity dispersion and X-ray distribution and temperature profile, \citet{Gomez12} reported the presence of an offset between the galaxy density peak and the peak of the X-ray emission and also a non-Gaussian galaxy velocity
distribution which supports a merger scenario for AS1063. \citet{Mohammed16} analyzed the power spectrum of the mass distribution and reported several morphological elongations, multiple clumps of mass, and a set of substructures that also indicate recent mergers in AS1063. Finally, \citet{Xie20} identified for both A370 and AS1063 radio haloes with size and energy %% consistent with the scaling relation between the cluster mass proxy and radio power
which supports the ongoing merging scenario for these clusters.

The distribution seen in the ICLf measured here shows the characteristic feature found for merging clusters, confirming by a completely independent method that both Abell 370 and Abell S1063 are merging systems. In addition, all of our ICLf measurements for the three filters in both clusters are also in agreement with the theoretical predictions for the total amount of light obtained by simulations of the ICL formation and evolution, since all of them are within the range of up to 40$\%$ for low-redshift clusters in the Universe \citep{Rudick11,Contini14}.

\section{Conclusions}

In this paper, we analyzed the ICLf in two FF clusters, Abell 370 and Abell S1063, in order to verify the correlation with the dynamical state reported by \citet{Jimenez18}. We have applied the same assumption-free algorithm to disentangle the ICL from other luminous sources in all three broadband filters F435W, F606W, and F814W in the original images \citep[CICLE,][]{Jimenez16}. The cluster membership was estimated using both PEAK and shifting gapper methods \citep{Owers2011} in the largest spectroscopic catalogs to the present date \citep[Rosati et al. in prep,][]{Karman15,Caminha16,Karman17,Lagattuta19}.

We have found ICLf values ranging between $\sim7\%-25\%$, and $\sim3\%-22\%$ at different wavelengths, for both A370 and AS1063, respectively. The associated errors are in a range of $\sim0.5\%$ to $\sim3\%$ and were estimated combining the photometric error, the cluster membership error, and the error in disentangling the ICL from the BCG. These measurements are in agreement with the range observed in merging clusters, and also present an excess in the ICLf for the intermediate filter, F606W. This excess can be explained by a greater presence of younger and/or of lower-metallicity stars in the ICL than in the cluster member galaxies, such as A- and F-type stars, already found by previous results derived from ICL colors analysis in the literature. The observed ICLf of AS1063 in the bluest filter is significantly lower compared to most of the clusters in the sample and this is in agreement with the previously found low contribution from young stellar populations when compared to the other clusters of the Frontier Fields program.  

The merging activity of both clusters is corroborated by different analyses in the literature, with results obtained through morphological, dynamic, and kinematic studies of the member galaxies as well as by the intracluster gas distribution and the presence of diffuse synchrotron emission. Our results therefore reinforce the ICL as a promising tool to determine the dynamical state of clusters of galaxies, and in addition, also confirms the CICLE as a robust and sophisticated technique to precisely measure the ICL contribution to the total amount of light in galactic systems.

\section*{Acknowledgements}

We thank the referee for constructive comments and suggestions that helped to improve the original manuscript. This work has received financial support from Fundação Carlos Chagas Filho de Amparo à Pesquisa do Estado do Rio de Janeiro (Faperj, Brazil) under the grant Nota 10 No. E-26/200.953/2019 (242872) and from Conselho Nacional de Desenvolvimento Científico e Tecnológico (CNPq, Brazil) under the grant No. 141631/2020-1. Y. J-T acknowledges financial support from the European Union’s Horizon 2020 research and innovation programme under the Marie Skłodowska-Curie grant agreement No 898633, and from the State Agency for Research of the Spanish MCIU through the "Center of Excellence Severo Ochoa" award to the Instituto de Astrofísica de Andalucía (SEV-2017-0709). R.A.D. acknowledges partial support support from NASA grants 80NSSC20P0540  and 80NSSC20P0597 and the CNPq grant 308105/2018-4. We also thank gratefully the computational support of Dr. Jailson Alcaniz. 

\section*{Data Availability}

The data underlying this article will be shared on reasonable request to the corresponding author.

%%%%%%%%%%%%%%%%%%%%%%%%%%%%%%%%%%%%%%%%%%%%%%%%%%



\bibliographystyle{mnras}
\bibliography{References} % if your bibtex file is called example.bib

\begin{thebibliography}{}
\makeatletter
\relax
\def\mn@urlcharsother{\let\do\@makeother \do\$\do\&\do\#\do\^\do\_\do\%\do\~}
\def\mn@doi{\begingroup\mn@urlcharsother \@ifnextchar [ {\mn@doi@}
  {\mn@doi@[]}}
\def\mn@doi@[#1]#2{\def\@tempa{#1}\ifx\@tempa\@empty \href
  {http://dx.doi.org/#2} {doi:#2}\else \href {http://dx.doi.org/#2} {#1}\fi
  \endgroup}
\def\mn@eprint#1#2{\mn@eprint@#1:#2::\@nil}
\def\mn@eprint@arXiv#1{\href {http://arxiv.org/abs/#1} {{\tt arXiv:#1}}}
\def\mn@eprint@dblp#1{\href {http://dblp.uni-trier.de/rec/bibtex/#1.xml}
  {dblp:#1}}
\def\mn@eprint@#1:#2:#3:#4\@nil{\def\@tempa {#1}\def\@tempb {#2}\def\@tempc
  {#3}\ifx \@tempc \@empty \let \@tempc \@tempb \let \@tempb \@tempa \fi \ifx
  \@tempb \@empty \def\@tempb {arXiv}\fi \@ifundefined
  {mn@eprint@\@tempb}{\@tempb:\@tempc}{\expandafter \expandafter \csname
  mn@eprint@\@tempb\endcsname \expandafter{\@tempc}}}

\bibitem[\protect\citeauthoryear{{Adami}, {Durret}, {Guennou}  \& {Da
  Rocha}}{{Adami} et~al.}{2013}]{Adami13}
{Adami} C.,  {Durret} F.,  {Guennou} L.,   {Da Rocha} C.,  2013, \mn@doi [\aap]
  {10.1051/0004-6361/201220282}, \href
  {https://ui.adsabs.harvard.edu/abs/2013A&A...551A..20A} {551, A20}

\bibitem[\protect\citeauthoryear{Allen, Schmidt, Ebeling, Fabian  \&
  Speybroeck}{Allen et~al.}{2004}]{Allen04}
Allen S.,  Schmidt R.,  Ebeling H.,  Fabian A.,   Speybroeck L.,  2004, \mn@doi
  [Monthly Notices of the Royal Astronomical Society]
  {10.1111/j.1365-2966.2004.08080.x}, 353

\bibitem[\protect\citeauthoryear{{Andersson}}{{Andersson}}{2006}]{Andersson06}
{Andersson} K.,  2006, {XMM observation of AS1063, one of the most luminous
  clusters known}, XMM-Newton Proposal

\bibitem[\protect\citeauthoryear{{Bertin}}{{Bertin}}{2011}]{Bertin11}
{Bertin} E.,  2011, in {Evans} I.~N.,  {Accomazzi} A.,  {Mink} D.~J.,   {Rots}
  A.~H.,  eds,  Astronomical Society of the Pacific Conference Series Vol. 442,
  Astronomical Data Analysis Software and Systems XX. p.~435

\bibitem[\protect\citeauthoryear{{Bertin} \& {Arnouts}}{{Bertin} \&
  {Arnouts}}{1996}]{BerAr1996}
{Bertin} E.,  {Arnouts} S.,  1996, \mn@doi [Astronomy and Astrophysics
  Supplement Series] {10.1051/aas:1996164}, \href
  {https://ui.adsabs.harvard.edu/abs/1996A&AS..117..393B} {117, 393}

\bibitem[\protect\citeauthoryear{Burke, Hilton  \& Collins}{Burke
  et~al.}{2015}]{Burke15}
Burke C.,  Hilton M.,   Collins C.,  2015, \mn@doi [Monthly Notices of the
  Royal Astronomical Society] {10.1093/mnras/stv450}, 449, 2353

\bibitem[\protect\citeauthoryear{{Caminha} et~al.,}{{Caminha}
  et~al.}{2016}]{Caminha16}
{Caminha} G.~B.,  et~al., 2016, \mn@doi [\aap] {10.1051/0004-6361/201527670},
  \href {https://ui.adsabs.harvard.edu/abs/2016A&A...587A..80C} {587, A80}

\bibitem[\protect\citeauthoryear{{Cenarro} et~al.,}{{Cenarro}
  et~al.}{2019}]{Cenarro19}
{Cenarro} A.~J.,  et~al., 2019, \mn@doi [\aap] {10.1051/0004-6361/201833036},
  \href {https://ui.adsabs.harvard.edu/abs/2019A&A...622A.176C} {622, A176}

\bibitem[\protect\citeauthoryear{{Connor} et~al.,}{{Connor}
  et~al.}{2017}]{Connor17}
{Connor} T.,  et~al., 2017, \mn@doi [\apj] {10.3847/1538-4357/aa8ad5}, \href
  {https://ui.adsabs.harvard.edu/abs/2017ApJ...848...37C} {848, 37}

\bibitem[\protect\citeauthoryear{{Conroy}, {Wechsler}  \& {Kravtsov}}{{Conroy}
  et~al.}{2007}]{Conroy07}
{Conroy} C.,  {Wechsler} R.~H.,   {Kravtsov} A.~V.,  2007, \mn@doi [\apj]
  {10.1086/521425}, \href
  {https://ui.adsabs.harvard.edu/abs/2007ApJ...668..826C} {668, 826}

\bibitem[\protect\citeauthoryear{{Contini}, {De Lucia}, {Villalobos}  \&
  {Borgani}}{{Contini} et~al.}{2014}]{Contini14}
{Contini} E.,  {De Lucia} G.,  {Villalobos} {\'A}.,   {Borgani} S.,  2014,
  \mn@doi [Monthly Notices of the Royal Astronomical Society]
  {10.1093/mnras/stt2174}, \href
  {https://ui.adsabs.harvard.edu/abs/2014MNRAS.437.3787C} {437, 3787}

\bibitem[\protect\citeauthoryear{De~Filippis, Sereno  \& Bautz}{De~Filippis
  et~al.}{2005}]{DeFilippis05}
De~Filippis E.,  Sereno M.,   Bautz M.,  2005, \mn@doi [Advances in Space
  Research] {10.1016/j.asr.2004.12.078}, 36

\bibitem[\protect\citeauthoryear{{DeMaio}, {Gonzalez}, {Zabludoff}, {Zaritsky}
  \& {Brada{\v{c}}}}{{DeMaio} et~al.}{2015}]{DeMaio15}
{DeMaio} T.,  {Gonzalez} A.~H.,  {Zabludoff} A.,  {Zaritsky} D.,
  {Brada{\v{c}}} M.,  2015, \mn@doi [\mnras] {10.1093/mnras/stv033}, \href
  {https://ui.adsabs.harvard.edu/abs/2015MNRAS.448.1162D} {448, 1162}

\bibitem[\protect\citeauthoryear{{DeMaio}, {Gonzalez}, {Zabludoff}, {Zaritsky},
  {Connor}, {Donahue}  \& {Mulchaey}}{{DeMaio} et~al.}{2018}]{DeMaio18}
{DeMaio} T.,  {Gonzalez} A.~H.,  {Zabludoff} A.,  {Zaritsky} D.,  {Connor} T.,
  {Donahue} M.,   {Mulchaey} J.~S.,  2018, \mn@doi [\mnras]
  {10.1093/mnras/stx2946}, \href
  {https://ui.adsabs.harvard.edu/abs/2018MNRAS.474.3009D} {474, 3009}

\bibitem[\protect\citeauthoryear{{Edwards}, {Alpert}, {Trierweiler}, {Abraham}
  \& {Beizer}}{{Edwards} et~al.}{2016}]{Edwards2016}
{Edwards} L.~O.~V.,  {Alpert} H.~S.,  {Trierweiler} I.~L.,  {Abraham} T.,
  {Beizer} V.~G.,  2016, \mn@doi [\mnras] {10.1093/mnras/stw1314}, \href
  {https://ui.adsabs.harvard.edu/abs/2016MNRAS.461..230E} {461, 230}

\bibitem[\protect\citeauthoryear{{Ellien}, {Durret}, {Adami}, {Martinet},
  {Lobo}  \& {Jauzac}}{{Ellien} et~al.}{2019}]{Ellien19}
{Ellien} A.,  {Durret} F.,  {Adami} C.,  {Martinet} N.,  {Lobo} C.,   {Jauzac}
  M.,  2019, \mn@doi [Astronomy and Astrophysics]
  {10.1051/0004-6361/201935673}, \href
  {https://ui.adsabs.harvard.edu/abs/2019A&A...628A..34E} {628, A34}

\bibitem[\protect\citeauthoryear{{Feldmeier}, {Ciardullo}, {Jacoby}  \&
  {Durrell}}{{Feldmeier} et~al.}{2004}]{Feldmeier04}
{Feldmeier} J.~J.,  {Ciardullo} R.,  {Jacoby} G.~H.,   {Durrell} P.~R.,  2004,
  \mn@doi [The Astrophysical Journal] {10.1086/424372}, \href
  {https://ui.adsabs.harvard.edu/abs/2004ApJ...615..196F} {615, 196}

\bibitem[\protect\citeauthoryear{{Furnell} et~al.,}{{Furnell}
  et~al.}{2021}]{Furnell21}
{Furnell} K.~E.,  et~al., 2021, \mn@doi [\mnras] {10.1093/mnras/stab065}, \href
  {https://ui.adsabs.harvard.edu/abs/2021MNRAS.502.2419F} {502, 2419}

\bibitem[\protect\citeauthoryear{{Giallongo} et~al.,}{{Giallongo}
  et~al.}{2014}]{Giallongo14}
{Giallongo} E.,  et~al., 2014, \mn@doi [\apj] {10.1088/0004-637X/781/1/24},
  \href {https://ui.adsabs.harvard.edu/abs/2014ApJ...781...24G} {781, 24}

\bibitem[\protect\citeauthoryear{{G{\'o}mez} et~al.,}{{G{\'o}mez}
  et~al.}{2012}]{Gomez12}
{G{\'o}mez} P.~L.,  et~al., 2012, \mn@doi [The Astronomical Journal]
  {10.1088/0004-6256/144/3/79}, \href
  {https://ui.adsabs.harvard.edu/abs/2012AJ....144...79G} {144, 79}

\bibitem[\protect\citeauthoryear{{Gonzalez}, {Zabludoff}  \&
  {Zaritsky}}{{Gonzalez} et~al.}{2005}]{Gonzalez05}
{Gonzalez} A.~H.,  {Zabludoff} A.~I.,   {Zaritsky} D.,  2005, \mn@doi [The
  Astrophysical Journal] {10.1086/425896}, \href
  {https://ui.adsabs.harvard.edu/abs/2005ApJ...618..195G} {618, 195}

\bibitem[\protect\citeauthoryear{{Guennou} et~al.,}{{Guennou}
  et~al.}{2012}]{Guennou12}
{Guennou} L.,  et~al., 2012, \mn@doi [Astronomy and Astrophysics]
  {10.1051/0004-6361/201117482}, \href
  {https://ui.adsabs.harvard.edu/abs/2012A&A...537A..64G} {537, A64}

\bibitem[\protect\citeauthoryear{{Gunn} et~al.,}{{Gunn}
  et~al.}{1998}]{Gunn1998}
{Gunn} J.~E.,  et~al., 1998, \mn@doi [\aj] {10.1086/300645}, \href
  {https://ui.adsabs.harvard.edu/abs/1998AJ....116.3040G} {116, 3040}

\bibitem[\protect\citeauthoryear{{Jim{\'e}nez-Teja} \&
  {Ben{\'\i}tez}}{{Jim{\'e}nez-Teja} \& {Ben{\'\i}tez}}{2012}]{Jimenez12}
{Jim{\'e}nez-Teja} Y.,  {Ben{\'\i}tez} N.,  2012, \mn@doi [The Astrophysical
  Journal] {10.1088/0004-637X/745/2/150}, \href
  {https://ui.adsabs.harvard.edu/abs/2012ApJ...745..150J} {745, 150}

\bibitem[\protect\citeauthoryear{{Jim{\'e}nez-Teja} \&
  {Dupke}}{{Jim{\'e}nez-Teja} \& {Dupke}}{2016}]{Jimenez16}
{Jim{\'e}nez-Teja} Y.,  {Dupke} R.,  2016, \mn@doi [The Astrophysical Journal]
  {10.3847/0004-637X/820/1/49}, \href
  {https://ui.adsabs.harvard.edu/abs/2016ApJ...820...49J} {820, 49}

\bibitem[\protect\citeauthoryear{Jiménez-Teja et~al.,}{Jiménez-Teja
  et~al.}{2018}]{Jimenez18}
Jiménez-Teja Y.,  et~al., 2018, \mn@doi [The Astrophysical Journal]
  {10.3847/1538-4357/aab70f}, 857

\bibitem[\protect\citeauthoryear{{Karman} et~al.,}{{Karman}
  et~al.}{2015}]{Karman15}
{Karman} W.,  et~al., 2015, \mn@doi [\aap] {10.1051/0004-6361/201424962}, \href
  {https://ui.adsabs.harvard.edu/abs/2015A&A...574A..11K} {574, A11}

\bibitem[\protect\citeauthoryear{{Karman} et~al.,}{{Karman}
  et~al.}{2017}]{Karman17}
{Karman} W.,  et~al., 2017, \mn@doi [\aap] {10.1051/0004-6361/201629055}, \href
  {https://ui.adsabs.harvard.edu/abs/2017A&A...599A..28K} {599, A28}

\bibitem[\protect\citeauthoryear{{Kneib}, {Mellier}, {Fort}  \&
  {Mathez}}{{Kneib} et~al.}{1993}]{Kneib1993}
{Kneib} J.~P.,  {Mellier} Y.,  {Fort} B.,   {Mathez} G.,  1993, Astronomy and
  Astrophysics, \href {https://ui.adsabs.harvard.edu/abs/1993A&A...273..367K}
  {273, 367}

\bibitem[\protect\citeauthoryear{{Krick} \& {Bernstein}}{{Krick} \&
  {Bernstein}}{2007}]{KrickBer07}
{Krick} J.~E.,  {Bernstein} R.~A.,  2007, \mn@doi [The Astronomical Journal]
  {10.1086/518787}, \href
  {https://ui.adsabs.harvard.edu/abs/2007AJ....134..466K} {134, 466}

\bibitem[\protect\citeauthoryear{{Lagattuta} et~al.,}{{Lagattuta}
  et~al.}{2019}]{Lagattuta19}
{Lagattuta} D.~J.,  et~al., 2019, \mn@doi [\mnras] {10.1093/mnras/stz620},
  \href {https://ui.adsabs.harvard.edu/abs/2019MNRAS.485.3738L} {485, 3738}

\bibitem[\protect\citeauthoryear{{Lah} et~al.,}{{Lah} et~al.}{2009}]{lah09}
{Lah} P.,  et~al., 2009, \mn@doi [\mnras] {10.1111/j.1365-2966.2009.15368.x},
  \href {https://ui.adsabs.harvard.edu/abs/2009MNRAS.399.1447L} {399, 1447}

\bibitem[\protect\citeauthoryear{Lin \& Mohr}{Lin \& Mohr}{2004}]{LinMohr04}
Lin Y.,  Mohr J.,  2004, \mn@doi [The Astrophysical Journal] {10.1086/425412},
  617

\bibitem[\protect\citeauthoryear{{Lotz} et~al.,}{{Lotz} et~al.}{2017}]{Lotz17}
{Lotz} J.~M.,  et~al., 2017, \mn@doi [The Astrophysical Journal Supplement
  Series] {10.3847/1538-4357/837/1/97}, \href
  {https://ui.adsabs.harvard.edu/abs/2017ApJ...837...97L} {837, 97}

\bibitem[\protect\citeauthoryear{{Medezinski}, {Broadhurst}, {Umetsu}, {Oguri},
  {Rephaeli}  \& {Ben{\'\i}tez}}{{Medezinski} et~al.}{2010}]{Mede10}
{Medezinski} E.,  {Broadhurst} T.,  {Umetsu} K.,  {Oguri} M.,  {Rephaeli} Y.,
  {Ben{\'\i}tez} N.,  2010, \mn@doi [Monthly Notices of the Royal Astronomical
  Society] {10.1111/j.1365-2966.2010.16491.x}, \href
  {https://ui.adsabs.harvard.edu/abs/2010MNRAS.405..257M} {405, 257}

\bibitem[\protect\citeauthoryear{{Melnick}, {Giraud}, {Toledo}, {Selman}  \&
  {Quintana}}{{Melnick} et~al.}{2012}]{Melnick12}
{Melnick} J.,  {Giraud} E.,  {Toledo} I.,  {Selman} F.,   {Quintana} H.,  2012,
  \mn@doi [Monthly Notices of the Royal Astronomical Society]
  {10.1111/j.1365-2966.2012.21924.x}, \href
  {https://ui.adsabs.harvard.edu/abs/2012MNRAS.427..850M} {427, 850}

\bibitem[\protect\citeauthoryear{{Merritt}}{{Merritt}}{1983}]{Merrit83}
{Merritt} D.,  1983, \mn@doi [The Astrophysical Journal] {10.1086/160571},
  \href {https://ui.adsabs.harvard.edu/abs/1983ApJ...264...24M} {264, 24}

\bibitem[\protect\citeauthoryear{{Mihos}}{{Mihos}}{2004}]{Mihos04}
{Mihos} J.~C.,  2004, in {Mulchaey} J.~S.,  {Dressler} A.,   {Oemler} A.,  eds,
  Clusters of Galaxies: Probes of Cosmological Structure and Galaxy Evolution.
  p.~277

\bibitem[\protect\citeauthoryear{{Mihos}, {Harding}, {Feldmeier}  \&
  {Morrison}}{{Mihos} et~al.}{2005}]{Mihos05}
{Mihos} J.~C.,  {Harding} P.,  {Feldmeier} J.,   {Morrison} H.,  2005, \mn@doi
  [The Astrophysical Journal Letters] {10.1086/497030}, \href
  {https://ui.adsabs.harvard.edu/abs/2005ApJ...631L..41M} {631, L41}

\bibitem[\protect\citeauthoryear{{Miller}}{{Miller}}{1983}]{Miller83}
{Miller} G.~E.,  1983, \mn@doi [The Astrophysical Journal] {10.1086/160974},
  \href {https://ui.adsabs.harvard.edu/abs/1983ApJ...268..495M} {268, 495}

\bibitem[\protect\citeauthoryear{{Miyazaki} et~al.,}{{Miyazaki}
  et~al.}{2002}]{Miyazaki2002}
{Miyazaki} S.,  et~al., 2002, \mn@doi [\pasj] {10.1093/pasj/54.6.833}, \href
  {https://ui.adsabs.harvard.edu/abs/2002PASJ...54..833M} {54, 833}

\bibitem[\protect\citeauthoryear{{Mohammed}, {Saha}, {Williams}, {Liesenborgs}
  \& {Sebesta}}{{Mohammed} et~al.}{2016}]{Mohammed16}
{Mohammed} I.,  {Saha} P.,  {Williams} L. L.~R.,  {Liesenborgs} J.,   {Sebesta}
  K.,  2016, \mn@doi [Monthly Notices of the Royal Astronomical Society]
  {10.1093/mnras/stw727}, \href
  {https://ui.adsabs.harvard.edu/abs/2016MNRAS.459.1698M} {459, 1698}

\bibitem[\protect\citeauthoryear{{Molnar}, {Ueda}  \& {Umetsu}}{{Molnar}
  et~al.}{2020}]{Molnar20}
{Molnar} S.~M.,  {Ueda} S.,   {Umetsu} K.,  2020, \mn@doi [\apj]
  {10.3847/1538-4357/abac53}, \href
  {https://ui.adsabs.harvard.edu/abs/2020ApJ...900..151M} {900, 151}

\bibitem[\protect\citeauthoryear{{Montes} \& {Trujillo}}{{Montes} \&
  {Trujillo}}{2014}]{MoTr14}
{Montes} M.,  {Trujillo} I.,  2014, \mn@doi [The Astrophysical Journal]
  {10.1088/0004-637X/794/2/137}, \href
  {https://ui.adsabs.harvard.edu/abs/2014ApJ...794..137M} {794, 137}

\bibitem[\protect\citeauthoryear{{Montes} \& {Trujillo}}{{Montes} \&
  {Trujillo}}{2018}]{MoTr18}
{Montes} M.,  {Trujillo} I.,  2018, \mn@doi [Monthly Notices of the Royal
  Astronomical Society] {10.1093/mnras/stx2847}, \href
  {https://ui.adsabs.harvard.edu/abs/2018MNRAS.474..917M} {474, 917}

\bibitem[\protect\citeauthoryear{{Montes} \& {Trujillo}}{{Montes} \&
  {Trujillo}}{2019}]{MoTr19}
{Montes} M.,  {Trujillo} I.,  2019, \mn@doi [\mnras] {10.1093/mnras/sty2858},
  \href {https://ui.adsabs.harvard.edu/abs/2019MNRAS.482.2838M} {482, 2838}

\bibitem[\protect\citeauthoryear{{Morandi}, {Ettori}  \&
  {Moscardini}}{{Morandi} et~al.}{2007}]{MoEtMos07}
{Morandi} A.,  {Ettori} S.,   {Moscardini} L.,  2007, \mn@doi [Monthly Notices
  of the Royal Astronomical Society] {10.1111/j.1365-2966.2007.11882.x}, \href
  {https://ui.adsabs.harvard.edu/abs/2007MNRAS.379..518M} {379, 518}

\bibitem[\protect\citeauthoryear{{Morishita}, {Abramson}, {Treu}, {Schmidt},
  {Vulcani}  \& {Wang}}{{Morishita} et~al.}{2017}]{Morishita17}
{Morishita} T.,  {Abramson} L.~E.,  {Treu} T.,  {Schmidt} K.~B.,  {Vulcani} B.,
    {Wang} X.,  2017, \mn@doi [The Astrophysical Journal]
  {10.3847/1538-4357/aa8403}, \href
  {https://ui.adsabs.harvard.edu/abs/2017ApJ...846..139M} {846, 139}

\bibitem[\protect\citeauthoryear{{Murante}, {Giovalli}, {Gerhard}, {Arnaboldi},
  {Borgani}  \& {Dolag}}{{Murante} et~al.}{2007}]{Murante07}
{Murante} G.,  {Giovalli} M.,  {Gerhard} O.,  {Arnaboldi} M.,  {Borgani} S.,
  {Dolag} K.,  2007, \mn@doi [Monthly Notices of the Royal Astronomical
  Society] {10.1111/j.1365-2966.2007.11568.x}, \href
  {https://ui.adsabs.harvard.edu/abs/2007MNRAS.377....2M} {377, 2}

\bibitem[\protect\citeauthoryear{{Oemler}, {Dressler}  \& {Butcher}}{{Oemler}
  et~al.}{1997}]{Oemler97}
{Oemler} Augustus J.,  {Dressler} A.,   {Butcher} H.~R.,  1997, \mn@doi [The
  Astrophysical Journal] {10.1086/303472}, \href
  {https://ui.adsabs.harvard.edu/abs/1997ApJ...474..561O} {474, 561}

\bibitem[\protect\citeauthoryear{{Owers}, {Randall}, {Nulsen}, {Couch}, {David}
   \& {Kempner}}{{Owers} et~al.}{2011}]{Owers2011}
{Owers} M.~S.,  {Randall} S.~W.,  {Nulsen} P. E.~J.,  {Couch} W.~J.,  {David}
  L.~P.,   {Kempner} J.~C.,  2011, \mn@doi [The Astrophysical Journal]
  {10.1088/0004-637X/728/1/27}, \href
  {https://ui.adsabs.harvard.edu/abs/2011ApJ...728...27O} {728, 27}

\bibitem[\protect\citeauthoryear{Patrikalakis \& Maekawa}{Patrikalakis \&
  Maekawa}{2010}]{PatriMae10}
Patrikalakis N.,  Maekawa T.,  2010, Shape interrogation for computer aided
  design and manufacturing, 1ed edn.
Springer

\bibitem[\protect\citeauthoryear{{Postman} et~al.,}{{Postman}
  et~al.}{2012}]{Post2012}
{Postman} M.,  et~al., 2012, \mn@doi [\apjs] {10.1088/0067-0049/199/2/25},
  \href {https://ui.adsabs.harvard.edu/abs/2012ApJS..199...25P} {199, 25}

\bibitem[\protect\citeauthoryear{{Puchwein}, {Springel}, {Sijacki}  \&
  {Dolag}}{{Puchwein} et~al.}{2010}]{Puch10}
{Puchwein} E.,  {Springel} V.,  {Sijacki} D.,   {Dolag} K.,  2010, \mn@doi
  [Monthly Notices of the Royal Astronomical Society]
  {10.1111/j.1365-2966.2010.16786.x}, \href
  {https://ui.adsabs.harvard.edu/abs/2010MNRAS.406..936P} {406, 936}

\bibitem[\protect\citeauthoryear{{Richard}, {Kneib}, {Limousin}, {Edge}  \&
  {Jullo}}{{Richard} et~al.}{2010}]{Richard10}
{Richard} J.,  {Kneib} J.~P.,  {Limousin} M.,  {Edge} A.,   {Jullo} E.,  2010,
  \mn@doi [Monthly Notices of the Royal Astronomical Society]
  {10.1111/j.1745-3933.2009.00796.x}, \href
  {https://ui.adsabs.harvard.edu/abs/2010MNRAS.402L..44R} {402, L44}

\bibitem[\protect\citeauthoryear{{Richstone}}{{Richstone}}{1976}]{Rich76}
{Richstone} D.~O.,  1976, \mn@doi [The Astrophysical Journal] {10.1086/154213},
  \href {https://ui.adsabs.harvard.edu/abs/1976ApJ...204..642R} {204, 642}

\bibitem[\protect\citeauthoryear{{Richstone} \& {Malumuth}}{{Richstone} \&
  {Malumuth}}{1983}]{RichMal83}
{Richstone} D.~O.,  {Malumuth} E.~M.,  1983, \mn@doi [The Astrophysical
  Journal] {10.1086/160926}, \href
  {https://ui.adsabs.harvard.edu/abs/1983ApJ...268...30R} {268, 30}

\bibitem[\protect\citeauthoryear{{Rudick}, {Mihos}  \& {McBride}}{{Rudick}
  et~al.}{2006}]{Rudick06}
{Rudick} C.~S.,  {Mihos} J.~C.,   {McBride} C.,  2006, \mn@doi [The
  Astrophysical Journal] {10.1086/506176}, \href
  {https://ui.adsabs.harvard.edu/abs/2006ApJ...648..936R} {648, 936}

\bibitem[\protect\citeauthoryear{{Rudick}, {Mihos}, {Frey}  \&
  {McBride}}{{Rudick} et~al.}{2009}]{Rudick09}
{Rudick} C.~S.,  {Mihos} J.~C.,  {Frey} L.~H.,   {McBride} C.~K.,  2009,
  \mn@doi [The Astrophysical Journal] {10.1088/0004-637X/699/2/1518}, \href
  {https://ui.adsabs.harvard.edu/abs/2009ApJ...699.1518R} {699, 1518}

\bibitem[\protect\citeauthoryear{{Rudick}, {Mihos}  \& {McBride}}{{Rudick}
  et~al.}{2011}]{Rudick11}
{Rudick} C.~S.,  {Mihos} J.~C.,   {McBride} C.~K.,  2011, \mn@doi [The
  Astrophysical Journal] {10.1088/0004-637X/732/1/48}, \href
  {https://ui.adsabs.harvard.edu/abs/2011ApJ...732...48R} {732, 48}

\bibitem[\protect\citeauthoryear{{Sartoris} et~al.,}{{Sartoris}
  et~al.}{2020}]{Sartoris20}
{Sartoris} B.,  et~al., 2020, \mn@doi [\aap] {10.1051/0004-6361/202037521},
  \href {https://ui.adsabs.harvard.edu/abs/2020A&A...637A..34S} {637, A34}

\bibitem[\protect\citeauthoryear{{Smail}, {Dressler}, {Kneib}, {Ellis},
  {Couch}, {Sharples}  \& {Oemler}}{{Smail} et~al.}{1996}]{Smail1996}
{Smail} I.,  {Dressler} A.,  {Kneib} J.-P.,  {Ellis} R.~S.,  {Couch} W.~J.,
  {Sharples} R.~M.,   {Oemler} Augustus J.,  1996, \mn@doi [The Astrophysical
  Journal] {10.1086/177799}, \href
  {https://ui.adsabs.harvard.edu/abs/1996ApJ...469..508S} {469, 508}

\bibitem[\protect\citeauthoryear{{Sommer-Larsen}, {Romeo}  \&
  {Portinari}}{{Sommer-Larsen} et~al.}{2005}]{SL2005}
{Sommer-Larsen} J.,  {Romeo} A.~D.,   {Portinari} L.,  2005, \mn@doi [\mnras]
  {10.1111/j.1365-2966.2005.08599.x}, \href
  {https://ui.adsabs.harvard.edu/abs/2005MNRAS.357..478S} {357, 478}

\bibitem[\protect\citeauthoryear{{Struble} \& {Rood}}{{Struble} \&
  {Rood}}{1999}]{StRo1999}
{Struble} M.~F.,  {Rood} H.~J.,  1999, \mn@doi [The Astrophysical Journal
  Supplement Series] {10.1086/313274}, \href
  {https://ui.adsabs.harvard.edu/abs/1999ApJS..125...35S} {125, 35}

\bibitem[\protect\citeauthoryear{{Sun}, {Donahue}  \& {Voit}}{{Sun}
  et~al.}{2007}]{Sun07}
{Sun} M.,  {Donahue} M.,   {Voit} G.~M.,  2007, \mn@doi [\apj]
  {10.1086/522690}, \href
  {https://ui.adsabs.harvard.edu/abs/2007ApJ...671..190S} {671, 190}

\bibitem[\protect\citeauthoryear{{Umetsu}, {Broadhurst}, {Zitrin}, {Medezinski}
   \& {Hsu}}{{Umetsu} et~al.}{2011}]{Umestu11}
{Umetsu} K.,  {Broadhurst} T.,  {Zitrin} A.,  {Medezinski} E.,   {Hsu} L.-Y.,
  2011, \mn@doi [The Astrophysical Journal] {10.1088/0004-637X/729/2/127},
  \href {https://ui.adsabs.harvard.edu/abs/2011ApJ...729..127U} {729, 127}

\bibitem[\protect\citeauthoryear{{Welch}, {McCandliss}  \& {Coe}}{{Welch}
  et~al.}{2020}]{Welch20}
{Welch} B.,  {McCandliss} S.,   {Coe} D.,  2020, \mn@doi [\aj]
  {10.3847/1538-3881/ab8ad8}, \href
  {https://ui.adsabs.harvard.edu/abs/2020AJ....159..269W} {159, 269}

\bibitem[\protect\citeauthoryear{{Williamson} et~al.,}{{Williamson}
  et~al.}{2011}]{Will11}
{Williamson} R.,  et~al., 2011, \mn@doi [The Astrophysical Journal]
  {10.1088/0004-637X/738/2/139}, \href
  {https://ui.adsabs.harvard.edu/abs/2011ApJ...738..139W} {738, 139}

\bibitem[\protect\citeauthoryear{{Xie} et~al.,}{{Xie} et~al.}{2020}]{Xie20}
{Xie} C.,  et~al., 2020, arXiv e-prints, \href
  {https://ui.adsabs.harvard.edu/abs/2020arXiv200104725X} {p. arXiv:2001.04725}

\bibitem[\protect\citeauthoryear{{Zwicky}}{{Zwicky}}{1951}]{Zwi51}
{Zwicky} F.,  1951, \mn@doi [Publications of the Astronomical Society of the
  Pacific] {10.1086/126318}, \href
  {https://ui.adsabs.harvard.edu/abs/1951PASP...63...61Z} {63, 61}

\makeatother
\end{thebibliography}

%\begin{thebibliography}{99}
%\end{thebibliography}

%%%%%%%%%%%%%%%%%%%%%%%%%%%%%%%%%%%%%%%%%%%%%%%%%%

%%%%%%%%%%%%%%%%%%%%%%%%%%%%%%%%%%%%%%%%%%%%%%%%%%

% Don't change these lines
\bsp	% typesetting comment
\label{lastpage}
\end{document}